\documentclass[]{aa}
\usepackage{graphicx}
\usepackage{natbib}

\begin{document} 

\title{One-armed oscillations in Be star discs}

\author{J.C.B. Papaloizou\inst{1} \and G.J. Savonije\inst{2}}

\institute{ Department of Applied Mathematics and Theoretical Physics,
Centre for Mathematical Sciences, Wilberforce Road, Cambridge CB3 0WA, UK 
\and Astronomical Institute `Anton Pannekoek', University of
  Amsterdam, Kruislaan 403, 1098 SJ Amsterdam, The Netherlands}

\date{Received ; accepted }

\abstract 
{}
{In this paper we  study  the effect of the quadrupole-term in the gravitational potential
 of a rotationally deformed central Be star on one armed density waves in the circumstellar disc.
 The aim  is to explain the observed long-term violet over red (V/R)
 intensity variations of the double peaked Balmer emission-lines,
 not only in cool Be star systems, but also in the hot systems like $\gamma$ Cas.} 
{We have carried out  semi-analytic  and numerical  studies of  low-frequency one armed global oscillations in near Keplerian discs around Be stars.
In  these we have  investigated surface density profiles for the circumstellar disc
which have inner narrow low surface density or  gap regions, just interior to  global maxima close to the rapidly rotating star, as well as the 
mode inner boundary conditions.}
{Our results indicate that  it is not necessary to invoke extra 
forces such as caused by line absorption from the stellar flux 
in order to explain the long-term V/R variations in the discs around massive Be stars.
When there exists a narrow gap between the star and its circumstellar disc, with the result
 that the radial velocity perturbation is non-zero at the inner disc boundary,
we find oscillation (and V/R) periods in the observed range for 
plausible  magnitudes for  the rotational quadrupole term.}
{}


\maketitle

\section{Introduction}
Many Be stars exhibit long-term periodicities in the relative 
intensity of the double peaked Balmer emission lines,
 the so-called V/R variations, which occur on timescales of typically
 a few years to a few decades, see e.g. \citet{Dachs87},
 \citet{Hub94} and \citet{Ok97} and references therein. 
\citet{Ok91} has suggested that a one armed (m=1) oscillation mode 
in the circumstellar disc about the Be star is responsible for the V/R variations. 
In a near Keplerian disc a one-armed mode shows up as a very slowly revolving 
perturbation pattern: the locally enhanced temperature and density causes stronger 
emission in one part of the disc. This results in either the
 violet- or red component of the double Balmer emission peaks to be brighter
 than normal when the compressed enhanced emission region is located 
in the approaching, respectively receding part of the disc. \citet{Ok91} applied 
a point mass potential for the Be star and obtained a slowly precessing global m=1 
pattern by the invoked pressure perturbations. 
However, in this way one generates {\it retrograde} modes which tend to 
propagate throughout the disc, while observations showed the V/R variations to occur 
only near the central star. 
\citet{Pap92} improved on this by considering a natural extension of this model 
  by including the quadrupole-term in the gravitational potential 
of the rapidly rotating and thus flattened Be star. 
Deviation from a $1/r$ point mass potential is known to give rise to precessing
 elliptic particle orbits. In a gaseous disc the deviation from a point mass potential 
leads naturally to slowly revolving {\em prograde} m=1 oscillation modes 
(if excited) which are confined to a region within a few stellar radii from the star. 
\citet{Pap92} showed that the propagation region extends in fact from the inner disc 
boundary up to about the point where the modal pattern speed equals the local particle 
precession speed. The radial size of this natural oscillation cavity is of order the stellar
 radius $R_s$ as the quadrupole component decays rapidly with increasing distance from the star.   

Meanwhile observations have shown that the perturbation pattern in several 
Be star-disc systems with V/R variations revolves indeed in a prograde sense, 
e.g. in $\beta^1$ Mon \citep{Telt94}. 
\cite{HV95} and \cite{HH97} calculated line profiles from discs with revolving m=1 modal 
patterns and showed that these eccentric modes can explain the observed variation of 
the asymmetric line profiles quite well. The existence of prograde m=1 modes close 
to the central star has also been directly detected by spectrally resolved 
interferometry of the Be stars $\zeta$ Tau \citep{Vak98} and $\gamma$ Cas \citep{Berio99}.

\cite{Ok97} argued that, although the observed density patterns appear 
indeed to revolve in the prograde sense, the proposed mechanism by \cite{Pap92} 
does not work for the hot early type Be star systems and he introduced a radiative 
force with the aim of confining the m=1 density waves to the inner part of these hot discs and induce prograde motions. 
Okazaki is indeed correct that in calculations with a rigid inner disc boundary \citep{SH93} unrealistically low disc temperatures had to be adopted. 
In a more recent study of m=1 modes \cite{FH06} note sensitivity of model predictions to disc model parameters, but comment that the predictive power of the model is better for the cooler Be stars where radiative forces are not added. In the latter work power law disc density profiles and rigid boundaries were also assumed.

However, in this paper we show, using both semi-analytic and numerical approaches, that if there is a narrow low  density region or gap between the star and its circumstellar disc, so that  the disc no longer has a rigid inner boundary but one that is essentially freely moveable,
and there is also a significant single surface density maximum not far away, 
prograde m=1 modes with periods of a few years  can be obtained even for hot Be star systems.
Although the mechanism by which a Be star ejects matter into the circumstellar disc 
is poorly understood, we note that for the mechanisms suggested in recent studies,
 such as non-radial oscillations in the rapidly rotating Be star \citep{Riv01,Sav05}, 
or magnetic effects, one may indeed expect such a gap region
to form when material with specific angular momentum larger than that corresponding 
to equatorial break up rotation is impulsively ejected from the surface during an outburst. 

The plan of the paper is as follows:
in Sect. 2 we give the basic equations for the oscillation modes
and consider them in the low frequency limit. In that limit 
we use variational principles to show that rigid wall conditions
are the least favourable for prograde modes and derive 
conditions on the stellar apsidal motion constant for their existence.
These are found explicitly for a class of surface density profiles, with
inner narrow  low surface density regions lying interior to nearby global surface density maxima that are also considered numerically. 
The semi-analytic and numerical approaches were found to be consistent.

In Sect.~3 we go on to apply the model to a hot   Be star: $\gamma$ Cas (B0.5 IVe)
and a cool Be star:  88 Her (B7 Ve). In both these cases prograde modes with reasonable
periods are found for an appropriate degree of rotatonal distortion and apsidal motion constant, without the need to introduce an extra radiation force. 
Finally in Sect.~4, we summarize our conclusions.

\section{Basic equations for m=1 density waves}
Let us introduce a non rotating cylindrical co\"{o}rdinate system $(r, \varphi,z)$ 
with its centre coinciding with that of the Be star. We assume 
the circumstellar discs about Be stars to be geometrically thin $(z <<r),$  
near Keplerian and also assume hydrostatic equilibrium in the z-direction ($v_z=0$). For simplicity we work with a two dimensional approximation by integrating over the z-direction. The z-averaged equations of motion  and the equation of continuity follow as
\begin{equation}
\frac{\partial{v_r}}{\partial{t}}+ v_r\, \frac{\partial{v_r}}{\partial{r}} + \frac{v_\varphi}{r}\,\frac{\partial{v_r}}{\partial{\varphi}} -\frac{v^2_\varphi}{r}= -\frac{1}{\Sigma}\, \frac{\partial{\Pi}}{\partial{r}}-\frac{\partial{\Psi}}{\partial{r}}
\label{Deq1}
\end{equation}

\begin{equation}
\frac{\partial{v_\varphi}}{\partial{t}} + v_r\, \frac{\partial{v_\varphi}}{\partial{r}} + \frac{v_\varphi}{r} \, \frac{\partial{v_\varphi}}{\partial{\varphi}} +\frac{v_r\,v_\varphi}{r} =-\frac{1}{\Sigma \, r}\frac{\partial{\Pi}}{\partial{\varphi}} - \frac{1}{r} \, \frac{\partial{\Psi}}{\partial{\varphi}}
\label{Deq2}
\end{equation}

\begin{equation}
\frac{\partial{\Sigma}}{\partial{t}} + \frac{1}{r}\, \frac{\partial{\left(\Sigma \, r\, v_r\right)}}{\partial{r}} + \frac{1}{r}\, \frac{\partial{\left(\Sigma \, v_\varphi\right)}}{\partial{\varphi}}=0
\label{Deq3}
\end{equation}

where we have introduced a vertically integrated surface density and pressure  
\begin{equation}
\Sigma =\int_{- \infty}^\infty \rho \, \rm{d}z \,\,\,\, \mbox{and} \,\,\,\, \Pi = \int_{- \infty}^\infty P \, \rm{d}z
\end{equation}
We adopt a polytropic (isothermal) equation of state $\Pi =~ K\, \Sigma^\gamma$ with $\gamma=1$ and define a (constant) sound speed $c=~\sqrt{\rm{d} \Pi / \rm{d} \Sigma}=\sqrt{K}$ in the disc.
 
Since the circumstellar discs in Be star systems have negilible mass, the gravitational potential $\Psi$ is that of the central Be star. It is essential to take the rotational deformation of the rapidly rotating Be star into account. Assuming uniform rotation and only weak distortions a  multipole expansion for $\Psi$ may be limited to the monopole and quadrupole terms \citep{Pap92}:
\begin{equation}
\Psi(r) = - \frac{G\, M_s}{r} - \frac{k_2\, \Omega_s^2\, R_s^5}{3\, r^3}
\label{pot}
\end{equation}
where $M_s$, $R_s$ and $\Omega_s$ are the mass, radius and angular rotation speed of the Be star and $k_2$ its apsidal motion constant \citep{Schw58}.

The unperturbed disc is assumed axisymmetric with negligible velocity in the radial direction, 
i.e. with velocity $\vec{v}(r)=~(0,\, \,r\, \Omega(r))$, whereby the angular velocity 
$\Omega(r)$ in the disc is given by  the condition for radial equilibrium 
\begin{equation}
r\, \Omega^2 = \frac{G \, M_s}{r^2} \, \left[1 + k_2\, f^2 \,\left(\frac{R_s}{r}\right)^2 \right] + \frac{1}{\Sigma} \,\frac{\rm{d} \Pi}{\rm{d} r} \label{eqilr}
\end{equation}
where we have used expansion (\ref{pot}) for the gravitational field and $f=\Omega_s/\Omega_c$ with $\Omega^2_c=G\, M_s/R_s^3$ as a measure for the stellar rotation speed.

\subsection{The linearised equations for the density waves}
After substituting the  Eulerian perturbations 
with harmonic time dependence  \\
 $\hat{\Sigma}^\prime (r,\varphi,t)=\Sigma^\prime(r) \,\,\rm{e}^{i\left(\sigma\, t- m\, \varphi \right)}$, 
\begin{equation}
\hat{v}^\prime_{\rm{r}} (r,\varphi,t)=v_{\rm{r}}^\prime(r) \,\,\rm{e}^{i\left(\sigma\, t- m\, \varphi\right)} \,\,\, ; \,\,\, \hat{v}^\prime_\varphi(r,\varphi,t)=v_\varphi^\prime(r) \,\, \rm{e}^{i\left(\sigma\, t- m\, \varphi\right)},
\end{equation}
which describe a density wave with m-fold azimuthal symmetry, in equations (\ref{Deq1})-(\ref{Deq3}) and neglecting all orders higher than linear we find, after applying  the operator relations $\frac{\partial{}}{\partial{t}} \rightarrow \rm{i}\, \sigma$  and $\frac{\partial{}}{\partial{\varphi}} \rightarrow - (\rm{i}\, m)$, the linearized oscillation equations  
\begin{equation}
\frac{\rm{d}W} {\rm{d}r}= -\rm{i} {\left(\bar{\sigma}^{2} -
\kappa^2\right) \over {\bar{\sigma}}}
 v^\prime_{\rm{r}} + \frac{2\, m \, \Omega}{\bar{\sigma}\, r} \,\, W 
\label{Leq1}
\end{equation}

\begin{equation}
\frac{\rm{d}v^\prime_{\rm{r}}} {\rm{d}r}= - \left(\frac{1}{r}
 + \frac{1}{\Sigma}\, \frac{\rm{d} \Sigma}{\rm{dr}} 
+ \frac{m}{\bar{\sigma}\, r^2} \, \frac{\rm{d}h}{\rm{d}r} \right) 
v^\prime_{\rm{r}} + \rm{i}\left(\frac{m^2}{\bar{\sigma}\, 
r^2}-\frac{\bar{\sigma}}{c^2}\right) W
\label{Leq2}
\end{equation}
where we have eliminated $v^\prime_\varphi$ and introduced
\begin{equation}
 W(r)=c^2\, \Sigma^\prime / \Sigma\,\,\,\, \mbox{and}\,\,\,\,  \bar{\sigma}=\sigma -m\, \Omega(r)
\label{defW}
\end{equation}
the oscillation frequency in a locally corotating frame.
The square of the epicyclic frequency is $\kappa^2 = (2\Omega/r) (dh/dr),$
with $h = r^2\Omega.$
The angular rate at which an orbit would precess in the absence of a perturbation to
the pressure force is given by
\begin{equation} \omega_p = \Omega - \kappa. \label{zero} \end{equation}
 From now on we adopt $m=1$ to obtain one-armed density waves.

\subsection{Low frequency limit}
Although equations (\ref{Leq1} - \ref{Leq2}) can be readily solved
numerically as they stand, we remark that 
the $m=1$ modes we consider are low frequency modes satisfying $|\sigma| \ll 
\Omega.$ This situation arises because the modes correspond
to elliptical distortions that precess on a long time scale compared
to the orbital period. In the low frequency limit the equations
may be reduced to a simplified form as in  \citet{Pap02} and  \citet{Pap05}. This enables general deductions
concerning the periods of oscillation and the way they are 
affected by boundary conditions to be made. To adapt equations (\ref{Leq1} - \ref{Leq2})
to this limit  we  assume that the system is of finite extent
and such that it can be characterised by a single dynamical time scale
$\Omega_0^{-1}$ that is everywhere significantly  shorter than the  precession time scale $\omega_p^{-1}$
and such that we can write $\Omega = O(\Omega_0).$
 We set $\omega_p = \epsilon \omega_{pr},$
where $\epsilon$ is a small parameter and $\omega_{pr}$
 is a relative
precession frequency scaled to be $O(\Omega_0).$  
We define a new eigenvalue $\lambda$ also of order $\Omega_0$
 by  setting $\sigma = \epsilon\lambda.$ In addition
we set  $W = \epsilon w $ and 
under the assumption that the disc is thin  $c^2 =\epsilon c_0^2.$

\noindent Equations (\ref{Leq1}) and (\ref{Leq2}) then give  to lowest order
in $\epsilon$
\begin{equation}
\frac{\rm{d}w} {\rm{d}r}= -2\rm{i} \left(\lambda -
\omega_{pr} \right) 
 v^\prime_{\rm{r}} - \frac{2w}{ r} 
\label{Leq1a}
\end{equation}
\and 
\begin{equation}
\frac{\rm{d}v^\prime_{\rm{r}}} {\rm{d}r}= - \left(\frac{1}{2r}
 + \frac{1}{\Sigma}\, \frac{\rm{d} \Sigma}{\rm{dr}}
\right)\,\,
v^\prime_{\rm{r}}+ \rm{i} 
\frac{w\Omega}{c_0^2}.
\label{Leq2a}
\end{equation}
Here we have used the fact that to within a correction of order $\epsilon,$
we have $h = r^2\Omega = \sqrt{GM_s r}.$
We can eliminate $w$ from equations (\ref{Leq1a}) and (\ref{Leq2a})
to obtain a single governing  second order differential equation
for $Q= \Sigma r^{1/2}v^\prime_{\rm{r}}$ in the form
\begin{equation}
{d \over dr}\left({r^{3/2}c_0^{2}\over \Sigma \Omega}\left({d Q\over d r}\right)\right)
={2r^{3/2}(\lambda - \omega_{pr})Q\over \Sigma }
\end{equation} 
or in terms of the original unscaled coordinates
\begin{equation}
{d \over dr}\left({r^{3/2}c^2\over \Sigma \Omega}\left({d Q\over d r}\right)\right)
={2r^{3/2}(\sigma - \omega_{p})Q\over \Sigma }\label{SLO}
\end{equation} 
Equation (\ref{SLO}) and its associated eigenvalue problem
 is of Sturm-Liouville form.
This enables us to make use of the associated
variational principle for the eigenfunctions
and eigenvalues  \citep{Cour53}.
We comment that it is possible to formulate this principle
using the product of $Q$ and an arbitrary function of $r$
as perturbation variable rather than $Q.$ However, the results
will be the same.

\subsection{Variational Principle}
In order to formulate a variational principle for
the eigenvalues, we must first consider the boundary conditions.
We consider the disc to have an inner boundary at $r=r_0$
and an outer boundary at $r=r_1.$ We also suppose that the surface density
is small but non zero at the boundaries as necessitated by our
adoption of an isothermal equation of state.
Then we may consider general boundary conditions for (\ref{SLO}) of the form
\begin{equation}{d Q \over dr} = A \, Q \label{BDRY}\end{equation}
At $r=r_0$ we take $A = A_0$ and at $r = r_1$ we take $A = A_1.$
At a rigid boundary, $Q=0,$
which can be thought of as  corresponding to $A \rightarrow \infty$
with $dQ/dr$ remaining finite. The boundary condition that the Lagrangian
variation of the pressure vanishes, as used in the numerical work below, 
corresponds to $A = (1/\Sigma)(d\Sigma /dr)$.

Recalling that for  Sturm-Liouville problems the eigenvalues are real 
and the eigenfunctions may be assumed to
be real without loss of generality (Courant \& Hilbert 1953), we derive 
an integral expression for the eigenvalue $\sigma$ from equation (\ref{SLO})
by multiplying it  by $Q$ and integrating  over $(r_0,r_1)$ in the form
\begin{equation}
\sigma = {I +
\left[{r^{3/2}c^2A\over 2\Sigma \Omega} Q^2\right]_1 -\left[{r^{3/2}c^2A\over 2\Sigma \Omega} Q^2\right]_0 
\over 
\int^{r_1}_{r_0}{r^{3/2}Q^2\over \Sigma }dr},\label{VARP}
\end{equation}
where
\begin{equation}I=\int^{r_1}_{r_0}\left({r^{3/2}\omega_{p}Q^2\over \Sigma } -
{r^{3/2}c^2\over 2\Sigma \Omega}\left({d Q\over d r}\right)^2\right)dr\end{equation}
and the subscript $0/1$ denotes evaluation at the inner/outer boundary respectively.
 Note that when the eigenfunction decays rapidly with distance the outer boundary
terms can become negligible and make no effective contribution. This is in fact
the case for the eigenfunctions we later consider. However, for completeness we shall retain these terms in the general analysis. 

We also recall from equations (\ref{eqilr}) and (\ref{zero}) that correct
to first order in $\epsilon,$
\begin {equation}
\omega_p = k_2 \Omega f^2 \left( {R_s\over r}\right)^2 
- {c^2\over 2\Omega r^2}{d\over dr}\left({r^2\over \Sigma}{d\Sigma \over dr}\right)
\equiv k_2\omega_{p,e}+ \omega_{p,c}\label{precfr} \end{equation}
and to adequate accuracy $\Omega$ may be replaced by the Keplerian value
$\Omega_K = (GM_s/r^3)^{1/2}$ in the above.

The expression (\ref{VARP}) forms the basis of the variational principle
for eigenvalues $\sigma.$ In particular, for finite $A_0$ and  $A_1,$  the largest eigenvalue, associated with
an eigenmode with no zeros, is an absolute
maximum with respect to variation with respect to arbitrary functions $Q.$
When a rigid wall condition is applied at a boundary, the same variational formalism 
applies, but trial functions nust be restricted to those that vanish at that boundary.

In this paper we are concerned with the question of the existence of prograde modes
which have $\sigma > 0.$ This issue is seen to be equivalent to that of whether the maximum of
(\ref{VARP}) is positive with respect to evaluation with all permissible trial functions.

\subsection{The existence of prograde modes and the dependence on boundary conditions}
We here point out that the existence of prograde modes is boundary condition sensitive
by firstly pointing out that there exists  sets of boundary conditions
for which the existence of prograde modes is assured. We go on to prove
that  a rigid wall boundary condition applied at either boundary minimizes the maximum
of (\ref{VARP}) and thus is the worst form of boundary condition for enabling
prograde modes. 

Consider the equation derived from (\ref{SLO}) when $k_2 =0$ and $\sigma =0.$
This is, with the help of (\ref{precfr}), found to be
\begin{equation}
{d \over dr}\left({r^3\over \Sigma }\left({d Q\over d r}\right)\right)
={Q r\over \Sigma }
{d\over dr}\left({r^2\over \Sigma}{d\Sigma \over dr}\right) \label{SLOP}
\end{equation}
In principle a solution of (\ref{SLOP}), which we denote by $Q = Q_0,$ can be used to define
values of $A_0$ and $A_1$ to use for the boundary conditions.
In fact $A_1$ can be specified and then $A_0$  determined
by inward integration from the outer boundary.
We comment that a specific example of a solution for a power law
$\Sigma \propto r^{-\gamma}$ is  $Q_0 \propto r^{-\delta},$
where $\delta = 1+\gamma /2 + \sqrt{1+\gamma^2/4}.$

The boundary conditions, or equivalently values of  $A_0$ and $A_1$ so determined, are such 
that when the  general eigenvalue problem is solved, there will be a prograde mode with
$\sigma > 0.$ This follows immediately from the fact that if the trial function $Q= Q_0$ 
is adopted, only the term proportional to $k_2$  survives in (\ref{VARP})
which must then be  positive. Hence  the maximum eigenvalue $\sigma$
must also be positive.

The fact that a rigid wall boundary condition always reduces a maximum of (\ref{VARP})
follows from the fact that adopting a rigid wall boundary condition through restricting trial functions to be such that $Q=0$ at one of the boundaries is equivalent to restricting the
search for a maximum to a subset of the possible trial functions that are available
when  a rigid wall condition is not used. In the latter case $Q$ is unrestricted at the boundary.
Thus, other things being equal, changing from a non rigid wall condition to a rigid wall
condition can only reduce the maximum eigenvalue $\sigma.$

\subsection{A condition on $k_2$ for the existence of a prograde mode\label{vrp}}
From the variational principle it follows that it is sufficient for a prograde
mode to exist that the numerator of (\ref{VARP}) be positive for some $Q$
which leads to

$$\hspace{-2cm} \int^{r_1}_{r_0}\left({r^{3/2}\omega_{p}Q^2\over \Sigma } -
{r^{3/2}c^2\over 2\Sigma \Omega}
\left({dQ \over dr}\right)^2\right)dr + $$
\begin{equation}\hspace{4mm} \left[{r^{3/2}c^2A\over 2\Sigma \Omega} Q^2\right]_1 -
\left[{r^{3/2}c^2A\over 2\Sigma \Omega} Q^2\right]_0 > 0. \end{equation}

Using (\ref{precfr}) this leads to
$$ \hspace{-2cm} k_2\int^{r_1}_{r_0}{r^{3/2}\omega_{p,e}Q^2\over \Sigma }dr \ge$$
$$\hspace{-1cm} \int^{r_1}_{r_0}{r^{3/2}\over \Sigma }\left(
{c^2\over 2\Omega}
\left({dQ \over dr}\right)^2- \omega_{p,c}Q^2\right)dr $$
\begin{equation}\hspace{4mm}- \left[{r^{3/2}c^2A\over 2\Sigma \Omega} Q^2\right]_1 +
\left[{r^{3/2}c^2A\over 2\Sigma \Omega} Q^2\right]_0 .\label{VARK} \end{equation}
This in turn leads to a variational principle  for the minimum value of $k_2$ 
for which a prograde mode exists. This is just the minimum that can be found from (\ref{VARK})
for any permissible $Q.$ Again because the possible trial functions for rigid wall
conditions are a subset of those allowed for other boundary conditions, the minimum
is larger in that case as is the corresponding value of $k_2.$
Accordingly free boundary conditions are always more favourable for prograde 
modes than are rigid wall conditions.

From general scaling arguments, we would expect that for surface densities sufficiently
peaked near $r = R_s,$ that this leads to a sufficient condition of the form
\begin{equation}k_2 f^2 > {\zeta c^2 R_s\over  G M_s} ,\label{con}\end{equation}
where $\zeta$ is a constant depending on the details of the surface density profile
and boundary conditions but which is larger  for rigid wall conditions relative to 
for example free boundary conditions. A condition of the same form
as (\ref{con}) was found in the case of a simple example presented below.

The variational principle for the minimum  $k_2$ for which prograde modes are possible
is equivalent to the eigenvalue problem obtained from equation (\ref{SLO})
when $\sigma$ is set to zero and $k_2$ is taken to be the eigenvalue. The minimum eigenvalue gives the
minimum value of $k_2$ for which prograde modes occur.

\subsection{ Density profiles that are highly  peaked near the inner edge}
We briefly consider the application of  (\ref{VARK}) to systems with
surface densities that are highly peaked near the inner edge.
To proceed we note that the trial function $Q =\Sigma$ satisfies the free
boundary condition (\ref{BDRY}). Using this and adopting a surface density
that is such that it and its radial derivative are very small at the boundaries so that the boundary terms become negligible, after some integrations by parts,
(\ref{VARK}) gives that  it is sufficient for prograde modes to exist that
\begin{equation} k_2\int^{r_1}_{r_0}\Sigma r^{3/2}\Omega f^2 \left( {R_s\over r}\right)^2 dr \ge
\int^{r_1}_{r_0}{\Sigma c^2\over \Omega r^{1/2}}dr 
.\label{VARK1} \end{equation}
 or after using  Kepler's law
 \begin{equation} \int^{r_1}_{r_0}\Sigma r^{3/2}\Omega\left( k_2f^2  {R_s^2\over r^2} 
 - {c^2 r\over GM }\right)dr \ge 0 
.\label{VARK2} \end{equation} 
From the above it is apparent that if $\Sigma$ is highly peaked, with a maximum 
at $r=R_{max}$ say, that prograde modes are guaranteed if
$k_2f^2 > c^2 R_{max}^3/( GM R_s^2 ).$ Thus for $R_{max}= 1.6R_s,$ we obtain
prograde modes if $k_2f^2 > 4.1(c^2 R_{s})/( GM  ).$
We now consider a specific example of a family of profiles and evaluate 
the condition  exactly. This turns out to be of similar form.
Thus, as has also been confirmed by normal mode calculations, the general form of our results does not depend on details of the surface density profile.

\subsection{A simple example}
We consider a surface density profile with a form very similar to
that adopted below for our numerical work, being given by

\begin{equation}
\Sigma(r)={\cal A} \, \left({R_s\over r}\right)^{\alpha}\exp{\left(\frac{-\beta R_s^3}{r^3}\right)}, 
\label{profilean}
\end{equation}
where ${\cal A}, \alpha,$ and $\beta$ are constants.

For this form of surface density, the equation for $Q,$ when $\sigma = 0$
is
\begin{equation}
{d^2 Q\over dx^2}+{dQ\over dx}\left({3+\alpha\over x}-{3\beta\over x^4}\right)
=Q\left({-\alpha \over x^2}-{2{\cal K}_2\over x^5}\right) \label{exx}\end{equation}
Here $x=r/R_s$ and ${\cal K}_2 = k_2f^2GM_s/( c^2 R_s)+3\beta.$
In terms of $z=x^{-3},$ we find
\begin{equation}
9z^2{d^2 Q\over dz^2}+{dQ\over dz}\left(3(1-\alpha)z +9{\beta z^2}\right)
=Q\left(-\alpha -2{\cal K}_2z\right) \label{exz}\end{equation}
Solutions of (\ref{exz}) can be found as power series expansions.
We adopt the solution that vanishes most rapidly as $z \rightarrow 0,$ 
or equivalently $x \rightarrow \infty$ which can be written, to within an arbitrary normalization factor,
 in terms
of a confluent hypergeometric function in the form \citep{Whitt63}
\begin{equation}
Q=z^q \ _1F_1( 2{\cal K}_2/( 9\beta) +q\,
 ;\, r/9+1\, ;\, -\beta z),
\end{equation}
where $r=3\sqrt{4+\alpha^2},$ and $q= (6+3\alpha +r)/18.$
Specification of an inner boundary condition, here taken at $r = R_s$ or $z=1,$
then determines ${\cal K}_2$ and hence $k_2f^2$ as an eigenvalue.
The smallest eigenvalue for fixed $\alpha$ and $\beta$ then leads to
the smallest value of $k_2$
for which prograde modes can exist.
For a rigid wall condition $Q=0$ and for vanishing Lagrangian change in pressure, 
$3zdQ/dz= Q(\alpha  - 3\beta),$ at $z=1.$

Once an inner boundary condition is adopted, it is a simple matter
to obtain the smallest value of $k_2f^2GM_s/(c^2 R_s)$ 
for which prograde modes are possible. Results for
various $\alpha$ and $\beta$ and different inner boundary 
conditions are given in table \ref{table1}.
These are fully consistent with the numerical work  we have carried out
(see below for further discussion).

\begin{table}
\caption{The values of the quadrupole factor $k_2 f^2$ above which there is a prograde $m=1$ mode in the disc determined from the semi-analytic theory as a function of the disc model parameters  $\alpha$ and $\beta$ defined in (\ref{profilean}), and the type of inner boundary condition.}
{\label{table1}}
\begin{tabular}{c c c c}
\hline \hline
Inner disc boundary  & $\alpha$  & $ \beta$ &  $k_2f^2GM_s/(c^2 R_s)$ \\
\hline
Free  &2  &7.5&4.7  \\
Free  &2   &6  &3.9  \\
Free  &2.5 &5.5&3.1  \\
Free  &2.5 &7.5&4.0  \\
Rigid &2.5 &0 &17.3  \\
\hline
\end{tabular}

\end{table}

\subsection{Strength of the quadrupole factor $k_2 f^2$ \label{s_aps}}
The structure and dynamical behaviour of the disc depends on the strength of the quadrupole term in equation (\ref{pot}) which is proportional to the factor $k_2 f^2$. Unfortunately, the apsidal motion constant $k_2$ and the stellar rotation speed factor $f=\Omega_s/\Omega_c$ are poorly determined stellar parameters. For B-type main sequence stars stellar evolution calculations yield  for $k_2$ typical values in the range $2.5 \times 10^{-3} - 10^{-2}$ \citep{Cl91,Cl95}. The rotation speeds are derived from line width measurements which includes the uncertain inclination factor $\sin{i}$ and the effect of rotational gravity darkening which is thought to have lead to too low estimates of the rotation speeds in earlier studies \citep{Por03} due to underweighting of the cooler, most rapidly rotating equatorial region. Yet it generally remains unclear how close a particular Be star is to critical rotation, although in some cases interferometry may be helpful \citep{Tyc04} and suggest large rotational deformation, e.g. for the Be star $\alpha$ Eri \citep{Sou03}. This has been questioned by \cite{Vin06}, who deduce a lower ratio of equatorial to polar radius for this star and a rotation rate $\Omega/\Omega_c \simeq 0.8$. 

\cite{Cran05} made a new detailed statistical study of the threshold rotation rates for the formation of Be star discs. He comes to the conclusion that the early type Be stars (O7e-B2e) exhibit a spread of equatorial rotation rates between a lower limit around $0.4-0.5 \,\Omega_c$ and an upper  limit near break up speed $\Omega_c$. He finds the spread in rotation rates to narrow progressively when considering later spectral types, approaching critical rotation speeds for the A0e stars. 

\cite{Fre05} studied the effects of fast rotation in early type stars and calculated synthetic spectra with classical plane-parallel non-LTE model atmospheres for rapidly rotating stars. They too find that gravity darkening can lead to serious underestimates of $V \sin{i}$ values when considering the rotational broadening of spectral lines. By comparing with spectra of 130 Be stars they conclude that the average angular rotation speed of Be stars is most probably $\Omega_s \simeq 0.9\, \Omega_c$. 
From all this we infer that for the hot, early type Be stars the quadrupole factor $k_2 f^2$ may lie somewhere between $4 \times 10^{-4}$ and $10^{-2}$, unfortunately quite a wide range.

\section{Numerical results}
\subsection{Adopted disc structure}
For the (poorly known) density distribution in the equilibrium disc we adopt the same form (\ref{profilean}) as used in the semi-analytic analysis above, except for an additional constant  $B$ which is chosen so as to provide a floor to the surface density such that it does not decrease to below $5\times 10^{-4}$ of its maximum value (and inhibit a solution for isothermal discs): 
\begin{equation}
\Sigma(r)=\frac{{\cal A}}{r^{\alpha}}\,\, \exp{\left(\frac{-\beta} {r^3}\right)}\,\, +\,\,  B 
\label{profile}
\end{equation}
 The exponential factor with the $\beta$ parameter is introduced to cut off the steeply increasing density profile near the star and to produce a gap (or low density zone) just outside the stellar surface. The formation of a gap seems a rather natural assumption for circumstellar discs that are generated by impulsive mass ejection from the central star. It allows us to apply a free boundary condition at the inner disc edge. As indicated above, the eigenvalue solution is only sensitive to the integrated disc properties and independent of the precise shape of the density maximum, as long as it is close to the star. Therefore simulations with an exponential cut off for different values of $\beta$ should be adequate to probe the dependence of the results to the disc's density profile near the star.

From here on we work in units for which $G=M_s=R_s=1$ and, as for the semi-analytic discussion, take $r=1$ for the inner disc boundary $r_0$, neglecting the small width of the gap between star and disc (the stellar radii being not well known anyhow).
Observationally derived (crude) power laws ($r^{-n}$) for the density distribution $\rho(r)$ in the circumstellar discs are typically $n\sim 2 - 4$ \citep{Wat87}. Adopting $n=3$ or 4 we find for a semi-Keplerian disc with disc semi-thickness (in z direction) $H\sim c/\Omega$ that $\alpha \simeq 3/2$ or $5/2$.

The profile of the surface density $\Sigma(r)$ for $\alpha=2$ and $\beta=7.5$ is plotted in Fig.~\ref{EF-gCas}. This disc extends quite far out from the star to $r\simeq 40$. The deviations from Keplerian rotation are strongest near the inner disc boundary where in our units $\Omega=1.0103$ (1.000) and the epicyclic frequency $\kappa=0.9881$. At $r=10$ the deviation from Keplerian rotation is already small with $\Omega=0.0308$ (0.0316) and $\kappa=0.0307$. The numbers quoted in brackets refer to the angular speed in a purely Keplerian disc (for which $\Omega=\kappa$).

We study the influence of the disc structure on the obtained oscillation periods by also considering  more compact discs with slightly different values of $\alpha$ and $\beta$. We assume the temperature in the isothermal discs is given by $T_d=2/3\, T_e$, where $T_e$ is the Be star's effective temperature, unless stated otherwise.

\subsection{Boundary conditions and numerical solution}
When doing numerical work we do not make
a low frequency approximation as above but solve equations (\ref{Leq1}-\ref{Leq2})
directly. 
However, as above, and unlike for previous calculations of m=1 modes in circumstellar discs \citep{SH93,Ok97} we consider the existence of a small low surface density 
region, or  gap, between the star and surrounding disc. Formation of a gap may be expected when the disc material was ejected from the star as a result of non-radial oscillations \citep{Riv01} or magnetic levitation with specific angular momentum larger than $\Omega_s \, R_s^2$.

We apply free boundary conditions to both the inner and outer boundary of the disc and require the Lagrangian pressure perturbation $\delta \Pi$ to vanish at both boundaries, or
\begin{equation}
v^\prime_{\rm{r}} =\frac{ \rm{i} \bar{\sigma}}{c^2} \, \frac{W}{\frac{\rm{d} \ln{\Sigma}}{\rm{d}r}}
\label{BC}
\end{equation}
unless specifically noted otherwise. These conditions correspond
to those used in the semi-analytic discussion.
We integrate equations (\ref{Leq1}-\ref{Leq2}) using a fourth order Runge-Kutta method with adaptive stepsize \citep{Press86} away from both disc boundaries, adopting an arbitrary boundary value $W$=1,  to a conveniently chosen fitting point in the central region of the disc. 
We then search for that (real) oscillation frequency $\sigma$ for which the difference in the ratio of $W/v^\prime_{\rm{r}}$ determined from integrations starting  on different sides
 of the fitting point becomes sufficiently small. The linear solution can then be scaled
 on each side  so that $W$ and $v^\prime_{\rm{r}}$ match separately.
 We comment that for the cases considered below, the location of the outer boundary was chosen such that further extension did not affect the results.

\subsection{A hot Be star: $\gamma$ Cas (B0.5 IVe)}
We now consider models for $\gamma$ Cas, the brightest Be star  in the northern sky 
for which we adopt $M_s=15 \rm{M}_\odot$, radius $R_s=~8 \rm{R}_\odot$ and effective temperature $T_e=25000\, \rm{K}$. \cite{FH06} adopted the same radius, but a larger mass ($18.3 \rm{M}_\odot$) and higher effective temperature ($T_e=33000 \,\rm{K}$) for this star.  We note that, according to the semi-analytic result in Sect.\ref{vrp}, these larger values require the quadrupole term $k_2 f^2$ to be larger by only a factor  $\propto c^2/M_s$, i.e. by a factor $\simeq 1.08$ in order to obtain the same oscillation period.  
The observed V/R quasi-periods of  $\gamma$ Cas fall in the range 4-7 years \citep{Doaz87}.

\begin{figure}[h]
{\resizebox{0.5\textwidth} {!} {\rotatebox{-90}{\includegraphics{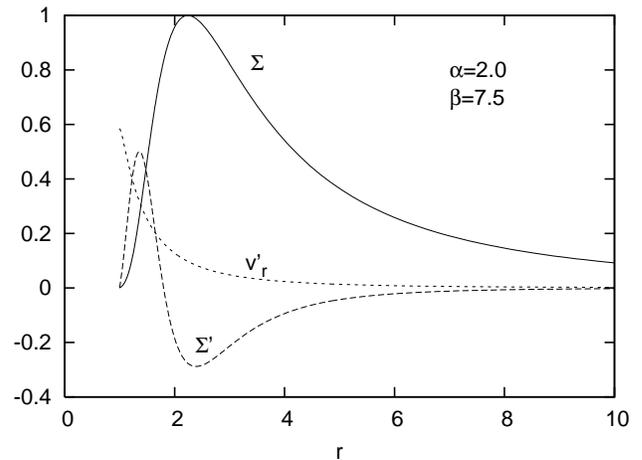}}}}
\caption{A model of $\gamma$ Cas with $M_s=15 \rm{M}_\odot$, $R_s=8 \rm{R}_\odot$ and $T_{e}=25000$K. The disc profile is defined by $\alpha=1.5$ and $\beta=7.5$ for which the disc has a free inner boundary (with a narrow gap between disc and star) as well as a free outer boundary near $r \sim 40$. Plotted are the density profile $\Sigma(r)$ of the unperturbed disc together with the eigenfunctions $\Sigma^\prime$ and $v^\prime_r$, in fact $\Im(v^\prime_r)$, for the m=1 mode with $P\simeq 7$ yr. We only show the inner 10 stellar radii from the stellar centre. The vertical scale is arbitrary.}
\label{EF-gCas}
\end{figure}

 We first consider the quite arbitrary semi-Keplerian circumstellar disc model described by (\ref{profile}) with $\alpha=2$ and $\beta=7.5$, for which the density profile $\Sigma(r)$ in the inner disc is shown in Fig.~\ref{EF-gCas}.
 Note that if we adopt a {\it rigid} inner boundary to the disc (i.e. adopt $\beta=0$, so that there is no gap) with $v^\prime_r=0$ there cannot exist prograde m=1 modes in the hot disc model considered here as the pressure forces dominate over the gravitational quadrupole effect and we obtain retrograde solutions that propagate all over the disc \citep{Pap92}. 

\begin{figure}[t]
{\resizebox{0.5\textwidth} {!} {\rotatebox{-90}{\includegraphics{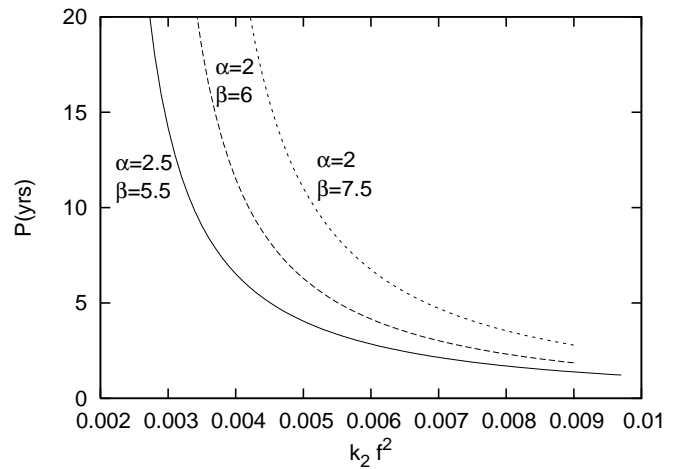}}}}
\caption{The oscillation (or V/R) periods in years versus the quadrupole factor $k_2 f^2$ of  (prograde) m=1 modes for the extended disc in the $\gamma$ Cas model with $\alpha=2$ and $\beta=7.5 $shown in Fig.~\ref{EF-gCas}, together with the results for the compact disc with $\alpha=2.5$ and $\beta=5.5$ shown in Fig.~\ref{EF-gCas2}. The observed V/R quasi-periods for $\gamma$ Cas range from 4-7 years. To see the effect of a steeper decline of $\Sigma$ towards the inner disc boundary, we have also plotted the results for $\beta=6$.}
\label{P-gCas}
\end{figure}

However, if we assume there is a (narrow) low-density region or gap between the star and the disc and consider a free inner disc boundary with a non-vanishing radial velocity of the gas, the situation changes drastically. Then the gas in the disc can oscillate in the radial direction right from the inner boundary where the gravitational quadrupole term is strongest, so that the latter term can dominate the dynamic behaviour. The numerical calculations show that a free inner boundary  leads to a very different outcome whereby we do find prograde m=1 oscillations in the observed period range, see Fig.~\ref{P-gCas}, consistent with our semi-analytic estimates. An observed V/R period of 7 years would for the disc model with $\alpha=2$ and $\beta=7.5$ thus require a rotation parameter  $k_2 f^2=0.0065$ which is rather large, although in the allowed range, see Sect.~\ref{s_aps}.
Fig.~\ref{EF-gCas} shows the shape of the eigenfunctions for a m=1 mode with period $P=6.6$ years. The shape of the eigenfunctions  varies little with $k_2 f^2$ (or period).  The shape of the density perturbation with a peak near $r=2.3$ is in fact rather similar to what \cite{Berio99} have inferred from their interferometric observations of $\gamma$ Cas: they deduce a prograde revolving m=1 pattern centred near $r\sim 2.5$ from the stellar centre.

\begin{figure}[h]
\resizebox{0.5\textwidth} {!} {\rotatebox{-90}{\includegraphics{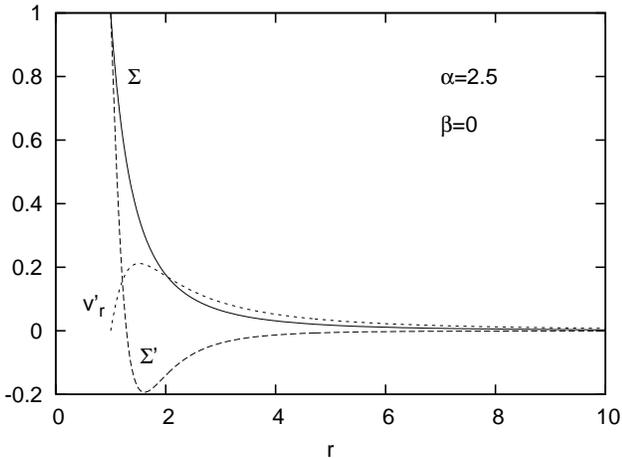}}}
\caption{A model of $\gamma$ Cas with a compact disc defined by $\alpha=2.5$ and $\beta=5.5$.
The circumstellar disc extends only to $r\sim 9$, with a steeper decay of $\Sigma$ towards the inner boundary in  comparison with the disc shown in Fig.~\ref{EF-gCas}.  Plotted are the density profile $\Sigma(r)$ of the unperturbed disc with the eigenfunctions $\Sigma^\prime$ and $v^\prime_r$ for the m=1 mode with period $P \simeq 4$ yr. The vertical scale is arbitrary.}
\label{EF-gCas2}
\end{figure}

For the extended disc shown in Fig.~\ref{EF-gCas} the also observed V/R period of 4 years would require a value of the quadrupole factor $k_2 f^2\sim 0.008$, closer to the upper limit of 0.01. However, the oscillation periods of the m=1 modes studied here appear sensitive to the size and adopted density profile of the disc.  An alternative profile $\Sigma(r)$ corresponding to $\alpha=2.5$ and $\beta=5.5$ is plotted in Fig.~\ref{EF-gCas2}, together with the m=1 eigenfunctions for $k_2 f^2=0.005$. The maximum for $\Sigma$ has shifted slightly more inward compared to the profile shown in Fig.~\ref{EF-gCas}, while the density vanishes already near $r\sim 9$.  Note that in this case the disc extends to only $r \simeq 9$, i.e. just beyond the propagation boundary of the m=1 mode near $r \simeq 6.5$.  For this more compact (measured by the location of the surface density maximum) disc the oscillation periods for a given quadrupole factor $k_2 f^2$ are significantly shorter and extend to smaller rotation values, as can be seen in Fig.~\ref{P-gCas}. For a compact (unevolved?) disc like this an m=1 oscillation period of 4 years requires a smaller quadrupole factor of $k_2 f^2 \simeq 0.005$.  The m=1 oscillation periods for an intermediate disc with $\alpha=2$ and $\beta=6$ are also plotted in Fig.~\ref{P-gCas}.

One can speculate that just after mass ejection and re-building of the circumstellar disc it may be rather compact and exhibits V/R quasi-periods in the lower part of the observed range while after some time, when angular momentum transport has taken place in the disc, it may get  smeared out leading to longer periods of the m=1 modes. This type of evolution would in Fig.~\ref{P-gCas} correspond to a vertical shift from the compact disc with the plotted lower branch ($\beta=5.5$) to the extended disc with the higher branch ($\beta=7.5$). That the two branches are not fully consistent in that they require slightly different quadrupole factors to explain the observed range in V/R periods should not surprise us in view of the arbitrariness of our disc models. It is interesting that the evolution from a 4 to 7 year V/R period occurred during the period 1970-1990 which coincided indeed with the build up of the disc around $\gamma$ Cas \citep{Doaz87}. 

It is also of interest to compare these results with those obtained using the semi-analytic theory. Values of $k_2 f^2\,GM_s/(c^2 R_s )$ above which prograde modes exist
for the values of $\alpha$ and $\beta$ used in  Fig.~\ref{P-gCas} are given in table
\ref{table1}. These can be converted to values of $k_2 f^2$ using 
$G M_s/(R_s c^2) = 1.5\times 10^{3}$ which is appropriate to this model. These values correspond
to the asymptotic limits for $P~\rightarrow~\infty$ in Fig.~\ref{P-gCas}.
They are found to be $0.0031,\,0.0026$ and $0.0021$ for the pairs
$(\alpha=2,\, \beta=7.5),$ $(\alpha=2,\, \beta=6)$ 
 and $(\alpha=2.5,\, \beta=5.5)$, respectively, and are accordingly consistent with
 the plots in Fig.~\ref{P-gCas}. 

\subsection{A cool Be star: 88 Her (B7 Ve)}
\begin{figure}[h]
\resizebox{0.5\textwidth} {!} {\rotatebox{-90}{\includegraphics{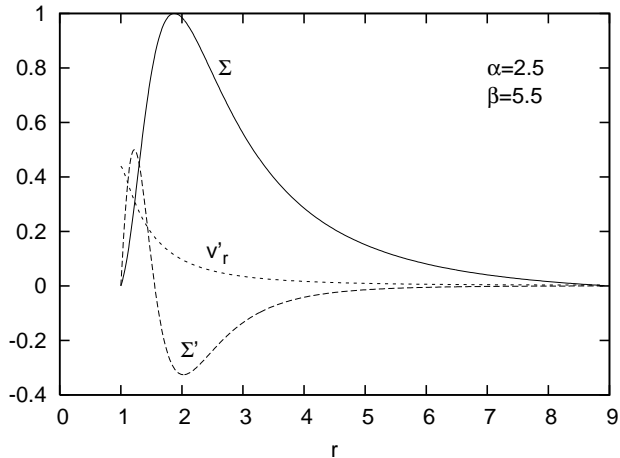}}}
\caption{A model of 88 Her with a disc defined by $\alpha=2.5$ and $\beta=0$ with $T_d=8333$ K. The disc touches the stellar surface at r=1 where it feels a rigid inner boundary ($v^\prime_r=0$), the free outer disc boundary is at $r \simeq 40$.  Plotted are the density profile $\Sigma(r)$ of the unperturbed disc with the eigenfunctions $\Sigma^\prime$ and $v^\prime_r$ for the m=1 mode with $P\simeq 15$ yr. The vertical scale is arbitrary.}
\label{EF-88Her2}
\end{figure}

\begin{figure}[t]
{\resizebox{0.5\textwidth} {!} {\rotatebox{-90}{\includegraphics{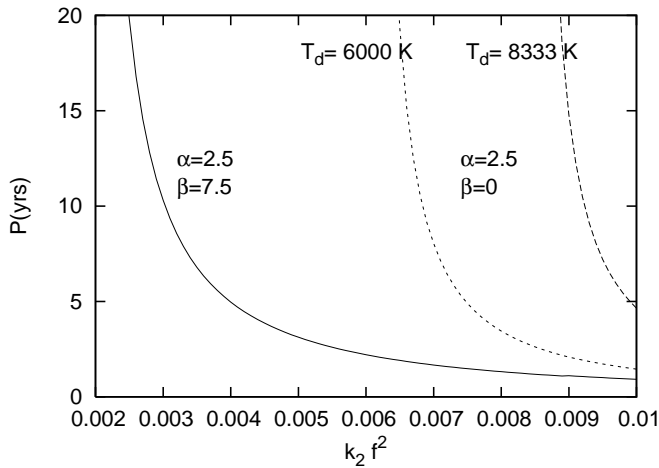}}}}
\caption{The oscillation (or V/R) periods in years versus  $k_2 f^2$ for the calculated (prograde) m=1 modes for a model of 88 Her, as indicated in the figure. The results for the two disc models (for two different disc temperatures) with $\beta=0$ correspond to $m=1$ modes with rigid wall inner boundary conditions.}
\label{P-88Her}
\end{figure}

We now turn to models for 88 Her, a Be star with estimated mass of $M_s=5\, \rm{M}_\odot$ with $R_s= 4\,\rm{R}_\odot$ and $T_e=~12500$\,K. For this cooler Be star system we consider two different discs, one with $\alpha=2.5$ and $\beta=7.5$ as before, i.e. with a gap between star and disc, whereby the disc extends out to $r\sim 19$, and another model with $\beta=0$, so that the inner disc touches the stellar surface. The boundary layer that exists in that case (assuming $\Omega_s < \Omega_c$) between the star and the inner disc is ignored here and replaced by the condition of a rigid wall: $v^\prime_r=0$.  
\begin{figure}[h]
\resizebox{0.5\textwidth} {!} {\rotatebox{-90}{\includegraphics{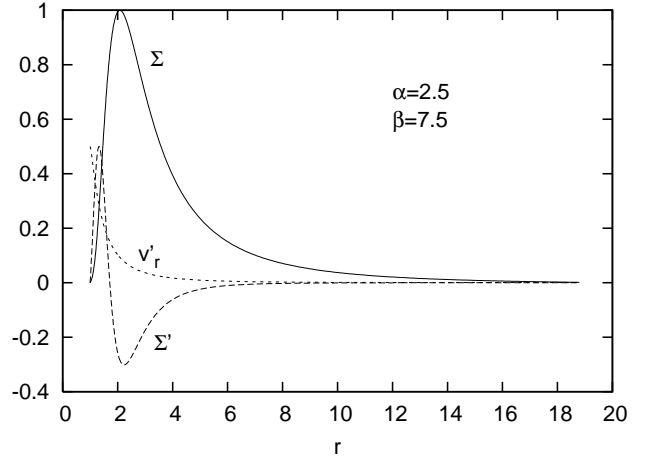}}}
\caption{A model of 88 Her  with $M_s=5\rm{M}_\odot$, $R_s= 4 \rm{R}_\odot$ and $T_e=12500$K and with a disc defined by $\alpha=2.5$ and $\beta=7.5$. The disc has a free inner boundary and extends to $r\simeq 19$. Plotted are the density profile $\Sigma(r)$ of the unperturbed disc with the eigenfunctions $\Sigma^\prime$ and $v^\prime_r$ for the m=1 mode with $P \simeq 3.3$ yr. The vertical scale is arbitrary.}
\label{EF-88Her}
\end{figure} 
The resulting disc profiles and calculated m=1 eigen modes are plotted in Fig.~\ref{EF-88Her} and \ref{EF-88Her2}, respectively. The periods of the calculated prograde m=1 modes are shown in Fig~\ref{P-88Her}. For the disc model that touches the stellar surface ($\beta=0$) we have calculated the m=1 modes for the standard disc temperature $T_d=2/3 \, T_e$ and also for a cooler disc with $T_d=6000$\,K. The disc model with free inner boundary ($\beta=7.5$) exhibits prograde m=1 modes with much lower oscillation periods and for significantly smaller values of the quadrupole factor $k_2\, f^2$ as compared to the discs with a rigid inner boundary.  
To compare these results with those obtained using the semi-analytic theory, we use the values of $k_2f^2\,GM_s/(c^2 R_s )$ above which prograde modes exist given in table
\ref{table1}. To obtain values of $k_2 f^2$ we take 
$GM_s/(R_s c^2) = 2\times 10^{3}$ which is appropriate to the model with
$(\alpha=2.5, \, \beta=7.5)$ and $G M_s/(R_sc^2) = 2.8\times 10^{3}$ for the model with $(\alpha=2.5,\, \beta=0)$ and $T_d = 6000$\,K. These values correspond
to the asymptotic limits for $P~\rightarrow~\infty$ in Fig.~\ref{P-88Her} and
 are found to be $0.002$ and $0.006$, respectively.
 These results are consistent with the data shown in Fig.~\ref{P-88Her}.
 In particular they confirm the much larger critical value
 of $k_2 f^2$  that applies in the case when a rigid wall inner boundary
 condition was applied to a distributed disc surface density profile.

\newpage
\section{Conclusions}
We have investigated the effect of the rotational flattening of the rapidly spinning Be stars on low-frequency global m=1 oscillations in their circumstellar discs
using both semi-analytic and numerical approaches.

We considered both a hot, early type Be star system like $\gamma$ Cas and a cool late type system like 88 Her and found that if it is assumed there exists a (narrow) low
surface density region or  gap between the star and the disc, such that the disc  inner boundary is free, low-frequency global m=1 oscillations can be supported with periods in the observed range of the long-term V/R variations of the Balmer emission lines for plausible values of the rotational factor $k_2 f^2(\le 0.008)$. Therefore it is not necessary to introduce radiation forces  \citep{Ok97} to obtain prograde m=1 modes that are contained in a oscillation cavity within a few stellar radii from a hot Be star surface. 

Our calculations suggest that the more compact circumstellar discs exhibit m=1 modes with smaller oscillation periods than the more extended discs for which the density maximum occurs slightly further out in the disc and for which the density decreases more slowly with increasing distance from the star.

Of course in its present  linear form our simple dynamical model cannot explain the full complexity of the observed variations in Be stars, in particular  non-linear effects need to  be included \citep{FH06}, especially in enigmatic systems like $\gamma$ Cas
\citep{Mot05}, but it appears to lead to a  natural explanation for the long term V/R variations in Be star systems, given the observational fact that there are circumstellar discs with slowly revolving m=1 patterns. The model naturally accounts for  the observed \citep{Berio99,Vak98} confinement and the prograde character of these patterns.
How the circumstellar discs are formed and how the low frequency one armed density waves therein are  excited still remains a puzzle. 

\bibliographystyle{aa}
\bibliography{sav}
\end{document}